\documentclass[10pt]{article}

\usepackage{amsmath}
\usepackage{amssymb}

\usepackage{graphicx}

\usepackage{cite}
\usepackage{subfigure}
\usepackage{color}

\usepackage[labelfont=bf,labelsep=period,justification=raggedright]{caption}

\makeatletter
\renewcommand{\@biblabel}[1]{\quad#1.}
\makeatother

\date{}

\newcommand{\model}{\begin{equation}
  \frac{dW(z,t)}{dt} =F_{\delta_{\rm{th}}}[I(z,t)] - r(z,t) + j(z,t)\label{eq:model}.
\end{equation}
}

\newcommand{\eqinput}{\begin{align}
I(z,t)&=\int_{{\rm
      Ex}}\mathrm{d}^{2}z'\;K[z-z'|W(z,t)]\;W^{*}(z',t),
\label{input}
\end{align}
}

\newcommand{\kernelfull}{\begin{align}
K[z-z'|W]&=K\left[ (z-z')\; {e} ^{-i\frac{\rm{arg}(W)}{2}} |1\right].
\label{kernel_full}
\end{align}
}

\newcommand{\eqkernel}{\begin{align}
K[z|1]=\left( \frac{z} {z^{*}} \right)^{2} \times 
\exp \Big\{-\frac{|z|^{2}} {2\,\sigma^{2}} - \mu\frac{|\rm{Im} (z)|} {[\rm{Re} (z)]^{2}}\Big\}.
\label{kernel}
\end{align}
}

\newcommand{\eqsuppression}{\begin{align} {r(z)}= \gamma_{{l}}W(z)+\gamma_{{g}}{H}(|W(z)|)
  \frac{W(z)}{|W(z)|}\text{\ensuremath{\intop}}_{{\rm
      In}}\mathrm{d}^{2}z^\prime|W(z^\prime)|.
\label{suppression}
\end{align}
}

\newcommand{\eqqmunu}{\begin{equation}
  Q_{\mu\nu} =\frac{1}{2}s
 \begin{pmatrix}
  -\sin(2\Theta)  & \cos(2\Theta)\\
 \cos(2\Theta) & \sin(2\Theta)
 \end{pmatrix}.
 \label{qmunu}
\end{equation}
}

\begin{document}

\begin{flushleft}
{\Large
\textbf{Director Field Model of the Primary Visual Cortex for Contour Detection}
}
\\
Vijay Singh$^{1}$, 
Martin  Tchernookov$^{1}$, 
Rebecca Butterfield$^{2}$,
Ilya  Nemenman$^{1,2\ast}$
\\
\bf{1} Department of Physics, Emory University, Atlanta, GA 30322, USA
\\
\bf{2} Department of Biology, Emory University, Atlanta, GA 30322, USA
\\
$\ast$ E-mail: Corresponding ilya.nemenman@emory.edu
\end{flushleft}

\section*{Abstract}
We aim to build the simplest possible model capable of detecting long,
noisy contours in a cluttered visual scene.  For this, we model the
neural dynamics in the primate primary visual cortex in terms of a
continuous director field that describes the average rate and the
average orientational preference of active neurons at a particular
point in the cortex.  We then use a linear-nonlinear dynamical model
with long range connectivity patterns to enforce long-range
statistical context present in the analyzed images. The resulting
model has substantially fewer degrees of freedom than traditional
models, and yet it can distinguish large contiguous objects from the
background clutter by suppressing the clutter and by filling-in
occluded elements of object contours. This results in high-precision,
high-recall detection of large objects in cluttered scenes.
Parenthetically, our model has a direct correspondence with the Landau
- de Gennes theory of nematic liquid crystal in two dimensions.
  
\section*{Introduction}
\begin{figure}[h]
\centerline{\includegraphics[scale=3]{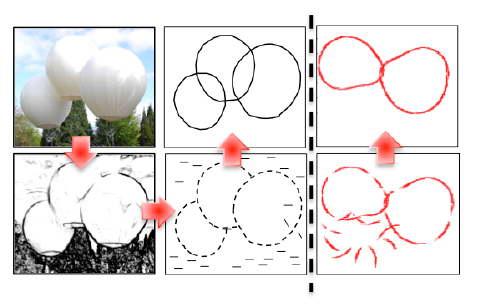}}
\caption{{\bf Contour Reconstruction Task:} A 2d image (left top; credit:
  {\tt `Pont de Singe', Olivier Grossetete. Photo: Thierry Bal}) is recorded as a field of
  contrast by the retina and the LGN (left bottom). V1 neurons respond
  to regions of contrast changes in a direction-selective manner,
  performing edge detection (middle bottom). The information from
  edges is integrated to reconstruct long contours (middle top). In
  this paper, we model the visual process starting from edges in V1;
  sample input (bottom) and output (top) to our model are on the
  right.\label{balloon}}
\end{figure}

To recognize an object in a visual scene, humans and other primates
process visual signals relayed through the retina
\cite{felleman1991distributed} in the ventral stream of the cortex.
Contour detection is a crucial part of this process
(Fig.~\ref{balloon}). It is carried out at early stages of the
processing in the primary visual
cortex (V1) of the brain \cite{creutzfeldt1978representation}. V1 consists of hundreds
of millions of neurons organized topographically into columns of
$\sim10^4\dots 10^5$ neurons each. Neurons in each column receive
inputs from a localized part of the visual field (called classical, or
feed-forward receptive field). They are directionally selective,
responding primarily to oriented edges within their receptive fields
\cite{hubel1968receptive,hubel1962receptive}.  Computational vision
models that account for such receptive fields of individual neurons
\cite{wallis1997invariant,itti1998model,lecun1998gradient,Thorpe:2001vn,boureau2010learning,Serre10042007}
typically incorporate them within feedforward hierarchical structures
similar to the cortex
\cite{riesenhuber1999hierarchical,riesenhuber2000models,Hinton:2006bg}. Such
feedforward models account for the visual processes on short time
scales, and achieve error rates as low as $\sim 10-20\%$ on
typical object detection tasks \cite{lecun2004learning,Serre10042007}.

It is believed that, {\em in vivo}, the error rate is reduced by
orders of magnitude by contextual information that influences local
processing, which may not be captured fully in such models
\cite{stettler2002lateral,angelucci2002circuits}. These collective,
recurrent dynamics span large spatiotemporal scales and are mediated
through thousands of axons laterally connecting distant columns
\cite{colonnier1981number}. These interactions are believed to
suppress the clutter present in the visual field, while simultaneously
binding edges into contours \cite{field1993contour}.

The goal of this paper is to build {\em the simplest model of the
  primary visual cortex} that simultaneously achieves two
contradictory tasks: clutter removal and occlusion filling. We do not
aim at the state of the art performance on complex natural images, but
rather ask what is the smallest set of computational primitives that
must be implemented in a model to achieve such detection and
integration of long contours in a nontrivial setting.  For this, we
focus on a proposal of a specific lateral connectivity among V1
neurons \cite{parent1989trace,gintautas2011model}, which incorporates
the Hebbian constraint that neurons that are excited simultaneously by
the same long, low-curvature contours should activate each other
\cite{parent1989trace}. However, in our model, we do not reproduce the
complexity of V1, which has $\sim100$ million neurons, with each
neuron having $\gtrsim10^3$ connections, some extending for many
millimeters. Instead, unlike most agent based discrete models, we
represent the activity as a {\em coarse-grained, continuous neural
  field}, which we model as a complex-valued field on the complex
plane, $W(z)$. The magnitude and the phase of $W$ represents the level
of excitation and the orientation of the dominant contour element at point $z$,
respectively. This coarse graining helps us to identify the minimal
features of the neural structure and dynamics that are essential for
contour recognition.

Importantly, our complex field approach is significantly simpler than
most other coarse-grained models, thus pushing the limits in identification
of the minimal set of the required computational primitives. Indeed,
typically the neural firing rate is represented as a real function of
three variables (position in the visual plane and the directional
sensitivity) \cite{zweck2004euclidean,bressloff2003functional}. In our
model, the firing rate is represented as a complex function of a
complex variable (or, equivalently, two real variables), which,
manifestly, has a lot fewer degrees of freedom. Previous approaches
that used a similar complex field representation
\cite{wolf1998spontaneous,wolf2003universality} have focused on
development, rather than on the visual performance of the cortex. Thus
it has been unclear if the simplified, lower-dimensional model can solve complex
visual tasks. Here we answer this question affirmatively.


\section*{Model}
We define the dynamical variables in our coarse-grained model as the
neural firing rate $s(x,y)$, $s\ge0$, over the two dimensional plane
$\mathcal{R}(x,y)$, and the orientation preference $\Theta$
of neurons, both averaged over a microscopic patch of the
cortex, which still contains many thousands of neurons. Such averaging
is traditional in, for example, fluid dynamics, where continuous
dynamics is sought from discrete agents. The neural activity is
invariant under parity (i.~e., an edge or its $\pi$ rotation results
in the same activity). Further, two equal edges at one point oriented
$\pi/2$ apart lead to cross orientation suppression, not forming a
dominant orientation at the point \cite{deangelis1992organization}.
Thus the fields $s$ and $\Theta$ are combined into a time varying
complex field $W(z,t)$ in a somewhat uncommon way, forming an object
called a {\em director} \cite{de1995physics}: $W(z,t)= s \times
e^{i2\Theta}$.  The magnitude of this field is the average firing
rate, and the argument is twice the average dominant orientation preference of neurons at a point $z=|z|e^{i\theta}=x+iy$
\cite{wolf1998spontaneous,wolf2003universality}.  We similarly
coarse-grain the input images, identifying the dominant orientation at
every point (see {\em Methods}). This orientation field serves as the
input to the model.  Note the crucial reduction in the number of
degrees of freedom in going from a more traditional description
$s=s(x,y, \Theta)$ to $W=W(z)$. One of the costs of the simplification
is the lost ability to represent multiple different orientations at
the same point, which happens when contours
intersect. Correspondingly, one of our goals is to verify that this
loss does not make it impossible to perform non-trivial visual tasks.

\begin{figure}[h]
\centerline{\includegraphics[scale=2]{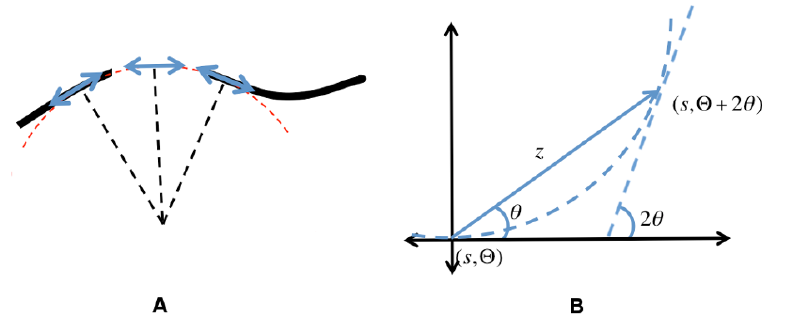}}
 \label{neuralresponse}
 \caption{ {\bf Co-circularity condition:} (A) Neurons send excitatory signals along approximately
   co-circular directions. Thus neurons in occluded gaps may get
   enough excitatory input along smooth contours to get excited
   without direct visual input. (B) The
   orientation at two points is said to be co-circular if they are
   tangential to the circle connecting the two points. If the
   orientation preference at the origin is along the real axis, the
   co-circular edge at a point $z=|z|e^{i\theta}$ has the orientation
   $2\theta$. Multiplication by $e^{i2\theta}$ can be written as:
   $e^{i2\theta}=(e^{i\theta})^2=\left( z/|z|\right)^2=\left(
     z/z^*\right)$.\label{occlusion}}
\end{figure}

Neurophysiological and psychophysical experiments
\cite{gilbert1989columnar,field1993contour,kovacs1993closed,stettler2002lateral,angelucci2002circuits}
and theoretical considerations \cite{parent1989trace} suggest that
neurons in V1 are laterally connected such that active neurons excite
nearby neurons with collinear or large-radius co-circular directional
preference. Conceptually, simultaneous input from several collinear or
co-circular neurons can excite other neurons that might otherwise not
be getting enough excitation from the visual field due to occlusion or
noise, cf.~Fig.~\ref{occlusion}{\bf A}. At the same time, neurons responding
to high spatial frequency clutter elements do not get sufficient
lateral excitation, and their activity decays. These collective
dynamics integrate information over large spatial scales.

We represent these phenomena in a traditional linear-nonlinear model,
where the neural field at a point $z$ is affected by a combination of
lateral synaptic inputs: \model Here $F_{\delta_{\rm{th}}}$ is some
sigmoidal function of the excitatory input $I(z,t)$, $r(z,t)$
describes the inhibitory contribution to the field, and $j(z,t)$ is
the stimulus.

The excitatory input, $I(z,t)$, combines synaptic input from all
points $z'$ in its interaction region `Ex' 
\eqinput 
where
$K[z-z'|W(z,t)]$ is the excitatory interaction kernel between the
fields at point $z'$ and $z$, when the field at $z$ is $W(z,t)$. The
kernel for an arbitrary orientation of $W(z)$ can be defined by an
appropriate rotation of the kernel defined for $W=1$ (parallel to the
real axis): \kernelfull

\begin{figure}[h]
\centerline{\includegraphics[scale=0.4]{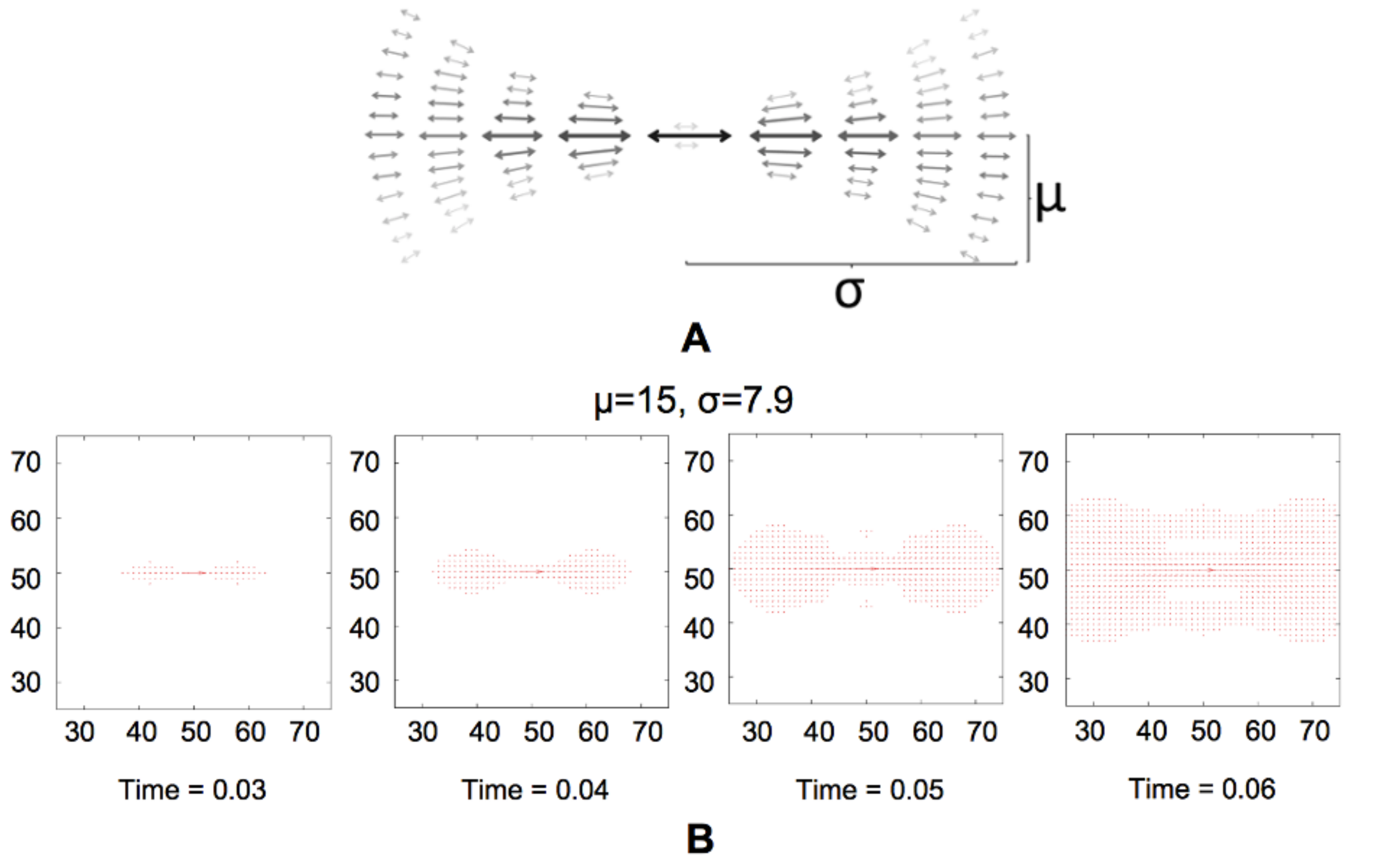}}
   \caption{{\bf Shape of the interaction kernel:} (A)Schematic shape of the interaction kernel $K[z-z'|1]$. Arrows
     represent the orientation preference and darkness and size
     represent the magnitude. (B) Results of dynamics with the kernel
     $K$ with the current $j(z,t)=\delta(z)\delta(t)$. Here, as
     everywhere in this work, we use $A = 5, \delta_{\rm{th}} = 5,
     \sigma =7.9, \mu = 15, \gamma_{\rm{g}} = 0.012, \gamma_{\rm{l}} =
     1$, which optimizes the performance according to a genetic
     algorithm search over the parameter space, see {\em Methods}.}
  \label{bowtie}
\end{figure}

Co-circular excitation may be represented as \eqkernel The first term,
derived in Fig.~\ref{occlusion}, determines the field direction at
$z$ that is co-circular to the field at $z'$. The $\sigma$ term in the
exponent determines the spatial range of the excitation. The $\mu$
term determines the smallest radius for which substantial co-circular
excitations still exist, giving the kernel and hence the induced
dynamics their characteristic bow-tie shapes \cite{parent1989trace},
see Fig.~\ref{bowtie}.  Note again the reduced complexity of this
model, where the kernel is defined by just two real-valued parameters,
instead of being inferred empirically from the data in a form of a
multi-dimensional matrix, as in Ref.~\cite{gintautas2011model} and
references therein.

We define the input nonlinearity using a complex step function:
\begin{equation}
\mathit{F_{\delta_{\rm{th}}}(I)}=\frac{I}{|I|}\times\;A\;H(|I|-\delta_{\rm{th}}),
\end{equation}
where $H$ is the Heaviside step function and $A$ determines the
maximum excitation strength. Smoother sigmoidal nonlinearities were
tried as well, but this had little effect on the results presented
below. If the total excitatory input is higher than the threshold
$\delta_{\rm{th}}$, then the field $W(z)$ gets a positive increment in
the direction of the total input. For this, the excitatory
contribution from a large part of the neural field must align in the
same direction, representing coincidence detection. While importance
of this coincidence detection phenomenon in vision is unclear, it is crucial in the context
of auditory signal processing \cite{jeffress1948place}. Thresholding
also suppresses clutter-induced spurious excitations, as it is
unlikely that the excitatory input from short clutter elements becomes
higher than the threshold in the absence of contextual support from
long contours.

The inhibition term $r$ represents two distinct phenomena: local
relaxation, which depends on the local field magnitude
\cite{dayanabbott}, and global inhibition \cite{Miconi:2010ei}, which
keeps the activity of the entire neural field in check (presumably
through intermediate inhibitory neurons, not modeled explicitly). In
the spirit of writing the simplest possible model, we represent
inhibition as linear, resulting in:
\eqsuppression Here $\gamma_l$ and $\gamma_g$ determine the rates of
local and global inhibition, and `In' stands for the range of global
inhibitory interactions. Combined with the non-linear excitation, this
linear inhibition produces bimodal asymptotic field values. Hence,
neurons can be defined as `active' or not.

\section*{Methods}

\subsection*{Image generation} 

Since our focus is not on practical image processing algorithms, we
focus on synthetic images in this work, as in
\cite{gintautas2011model}. This makes it easier to analyze effects of
various image properties on the performance.

{\em\bf Targets --} The ``amoeba'' objects (long closed contours with
gaps) are generated by choosing a center at a random point in the
image, and then drawing the amoeba around this point in polar
coordinates, with the radius as a superposition of periodic functions
with different radial frequencies,
$\rho(\phi)=\Sigma_{{k}=0}^{n}\;a_{{k}} \sin (k \phi + \phi_k) $. The
Fourier coefficients $a_k$ are generated randomly from a normal
distribution ($\sigma = 1$), with $k\le n=3$, and the phases $\phi_k$
are uniformly distributed between 0 and $2\pi$. To create amoebas that
are about the same size, the coefficients are further constrained such
that the minimum and the maximum radii of the resulting amoeba and
their ratios obey $0.2L<R_{\rm min}<R_{\rm
  max}<0.3L,\;0.4<\frac{R_{\rm min}}{R_{\rm max}}<0.6$, where $L$ is
the image size. The input current then is $j(z)=\delta(z-z_e)e^{i2\Theta}$, for
every point $z_e$ within 1 lattice spacing away from any point on the
amoeba contour, where $\Theta$ is tangential to the contour at that
point.  While generating an amoeba, we also determine an exclusion
region around it of 8 lattice sites. Clutter elements (see below) with
orientations parallel to the closest amoeba segment are not allowed in
these regions. Without such exclusion, a nearby clutter edge could
help amoeba detection, which would artificially elevate the measured
performance. We prefer to err on the side of underestimating the
performance, and hence we remove these ambiguous cases.

{\em\bf Occlusions --} We simulate occlusions and noise in real-world
images by removing parts of amoebas. A random number of 2-4
segments with random angular length combining to the total of $\sim
25\%$ of the amoeba length are chosen at random positions along the
amoeba contour. Within the chosen segments, the input current $j(z)$
is then set to zero.

{\em\bf Clutter -- }We need the clutter to be indistinguishable from the
targets by curvature, brightness, and other local statistics, so that
object detection is impossible without long-range contextual
contour integration afforded by co-circular connectivity.  Thus clutter is
generated by first generating an amoeba as described above,
partitioning it into segments, and then randomly shuffling and
rotating the segments to break long-range contour continuity.
Specifically, the model cortex is divided into $5\times5$ square
regions, which are then randomly permuted. The center-of-mass (CoM) of
an image within each region is computed, and the dominant angular
orientation is determined. Then each region is rotated around its CoM
by a random angle, subject to a constraint that the resulting dominant
orientations of neighboring regions are different. The constraint
ensures that the clutter does not form long range target-like
structures.

{\em\bf Combined images --} One or two targets and clutter resulting from
breakup of one or two additional targets were then superimposed
together to form test images, see Fig.~4, for an
example. Clutter in the exclusion zones along the amoeba contours was
then removed, as described above.

{\em\bf Transforming pixel images --} Images used previously in psychophysics
experiment (Fig.~4) were imported into MATLAB and then
converted to grey scale using {\tt rgb2grey}. The resulting matrix was
then thresholded and converted into a binary matrix. A 2D Gabor filter
was used to find edges in this bitmap image. For each point in the
image, we find the convolution of a Gabor filter $(\sigma_{\rm
  smaller}=10\;\rm{pixel}, \; \sigma_{\rm longer}=100\; \rm{pixel}, \;
\rm{convolution\;range}= 20\; \rm{pixel} \times 20\; \rm{pixel})$ with
the image at $(360/n)$ angles where $n=100$. The direction with the
maximum convolution is taken as the orientation of the visual field at
the point, and the result of the convolution as the field
magnitude. The image thus processed is presented as an input for
simulations.

\subsection*{Simulations}
The time evolution of the model is studied on a square lattice of a
linear size $L=100$ with periodic boundary conditions using Euler
iteration method. The lattice discretization is done for simulation
purposes, and should not be viewed as a representation of discrete
neurons; we are not aware of numerical algorithms able to simulated
our model dynamics without discretizing the space first.

In each iteration cycle we first calculate the total input $I$ at each
point $z$ from all other points $z'$ in the excitation region `Ex'
using a precomputed interaction kernel $K[z-z'|1]$ on a $4 L\times4L$
{\em kernel lattice}. Square discretization destroys the angular
symmetry of the kernel evaluated at an arbitrary $z$. The following
procedure restored the symmetry.  First, to calculate the contribution
from $z'$ to $I(z)$, the kernel lattice is superimposed on the image
lattice with the origin of the kernel lattice at point $z$ of the
image lattice. Next the kernel lattice is rotated by $\frac{{\rm
    arg}({W}(z'))}{2}$ with respect to the image. Then the
contribution from the point $z'$ to $I(z)$ is $W^*(z') \times
K(0,z'')$, where $z''$ is the point on kernel lattice closest to
$z'$. The total input $I(z)$ is then the sum of contributions from all
points $z'$ in the excitatory interaction region `Ex'.  After the
input is calculated, if $|{I(z)}|\;>\;\delta_{\rm{th}}$, then the
field is incremented $W(z,t+\Delta t) \leftarrow
W(z,t)+{A}\frac{{I(z)}}{{|I(z)|}}\Delta t$, where $\Delta t$ is the
time step. To account for degradation, we finally set $W(z,t+\Delta t)
\leftarrow W(z,t+\Delta t)\times \exp[-{r}(z) \times\Delta t /
W(z,t+\Delta t)]$, where ${r}(z)$ is as in Eq.~(5). 
To the first order in $\Delta t$, this is equivalent to the dynamics
in Eq.~(1). 
However, this exponential form removes the large fluctuations in
$r(z)$ when $W(z)\approx 0$.

In our simulations, the excitation range $`\rm{Ex}'$ is $3\sigma$,
where $\sigma$ is the effective spatial range of the kernel
$K[z-z'|W(z)]$. For global inhibition range $`\rm{In}'$ is the entire
lattice. The model is easily modified to restrict the suppression to a
smaller inhibition region. 

We first chose the parameter $\mu$ to be similar to the curvature of a
typical amoeba. Next $\sigma$ was chosen such that it was larger than the typical extent
of the occluded amoeba segments. The initial values of $\gamma_{\rm{l}}$ and
$\gamma_{\rm{g}}$ were determined using steady state
analysis of the model, which leads to $(N
\gamma_{\rm{g}}+\gamma_{\rm{l}})W_{0} \sim F_{\delta}(\infty)$, where
$N$ is the typical number of points with non zero field, and $F$
is the thresholding function as defined in Eq.~(1). 
Setting $\gamma_{\rm{l}}=1$ and $W_0=1$, we thus constrain all other
parameters. Using these initial values, some coarse parameter optimization was done by simply observing the simulations while the parameters were varied. After that genetic algorithm was
used to optimize the model for maximum simultaneous precision and
recall (see Results for definitions). We used the area under the precision-recall curve as our fitness function. Parameters were changed by a percentage drawn from a uniform distribution (from -1\% to 1\%) and the fitness function was recalculated for the new parameters. Then, the new parameters were either accepted or rejected according to whether $1/[1+\exp(new~area - old~area)/0.005] >$ random variable drawn from uniform distribution on (0,1). The parameter 0.005 acts as the temperature. The final optimized values of the parameters used for
simulations presented here were: $A = 5, \delta_{\rm{th}} = 5, \sigma
=7.9, \mu = 15, \gamma_{\rm{g}} = 0.012, \gamma_{\rm{l}} = 1$.

The code was implemented in C, compiled with the gcc v.~4.7, and
optimized with OpenMP libraries. Simulations were performed on a
computer with Intel i7 2600k (clock speed 3.4 GHz). The simulation
time for 250 iteration cycles for one image took about 10s.  All model
dynamics times were measured in units of $1/\gamma_{\rm{l}}$, which
was set to 1 in our simulations.

\section*{Results}
\begin{figure}[h]
\centerline{\includegraphics[scale=1.3]{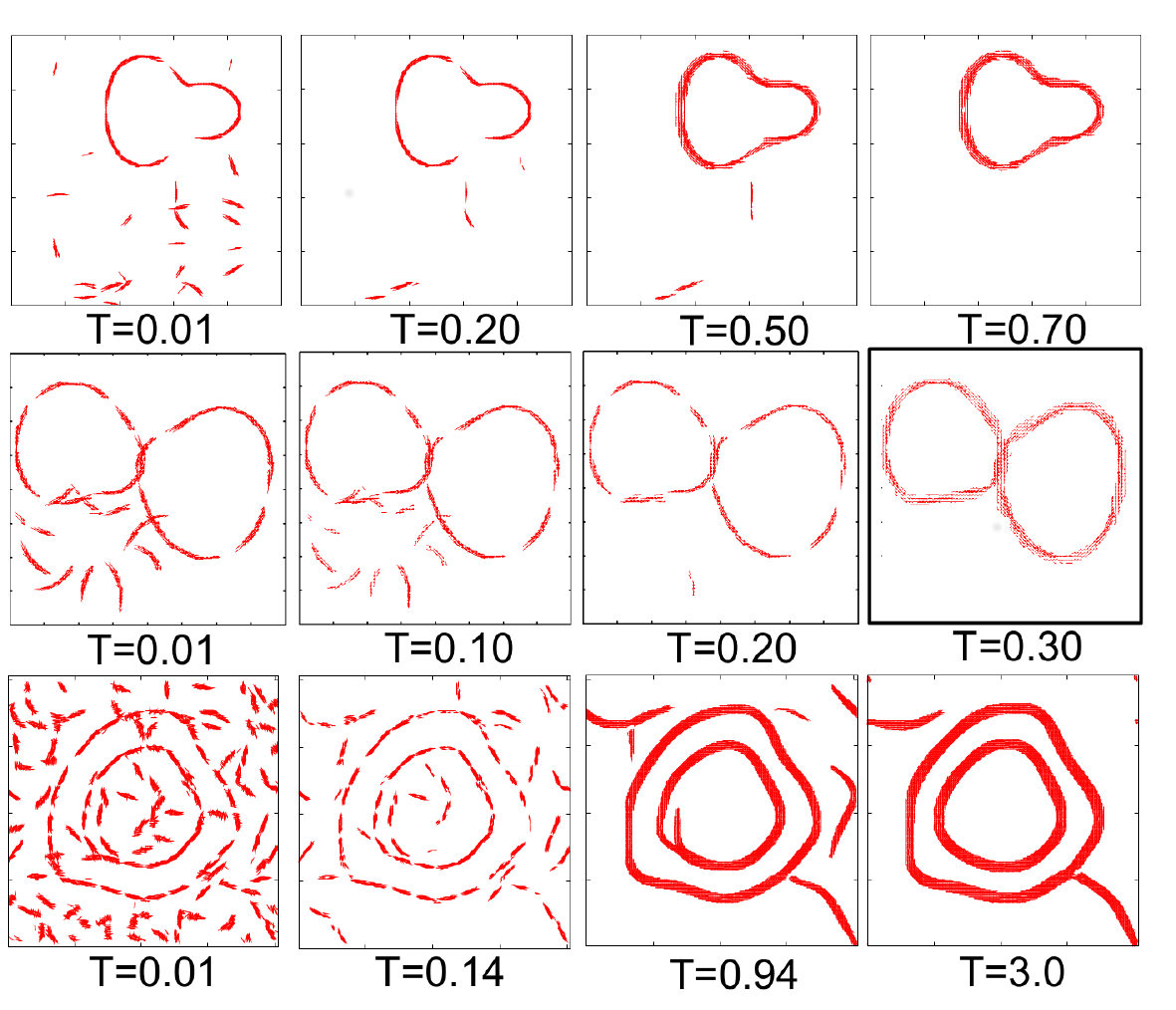}}
\caption{{\bf Neural field dynamics:} (top and middle) Time evolution of the neural field for sample
  images. The magnitude (line width) and the direction of the field are
  plotted at every point where the strength of the field is higher
  than a cutoff (0.35). The parameters of the dynamics are as in
  Fig.~\ref{bowtie}. Dynamics removes the clutter and fills in the
  occlusion gaps. However, spurious activity (widening lines) appears
  for large simulation times, so that the best performance is obtained
  for intermediate times. (bottom) Performance of the model on an
  image used in psychophysics experiments
  \cite{altmann2003perceptual}. Like human subjects, the model can
  identify, complete, and bind together long punctuated contours. }
 \label{timeevolution}
\end{figure}

Figure \ref{timeevolution} (top and middle) shows the time evolution of the neural
field $W(z,t)$ in our coarse-grained model for a sample input image,
generated as described in {\em Methods}, where a large contiguous
contour with gaps (an amoeba) is superimposed on clutter. The gaps
model occlusion of contours by other objects and noise in the earlier
stages of visual processing. Similarly, Figure \ref{timeevolution} (bottom) illustrates the
model output for an image previously used in psychophysics experiments
with human subjects \cite{altmann2003perceptual}. Its simplicity
notwithstanding, the model performs qualitatively similar to humans in
that long contours implied by collinearity of nearby edge segments are
easily detected. The gaps in amoeba targets get filled, while the
clutter decays with time, resulting in emergence of long
contours. Note also that spurious activity appears around contours at
large simulation times. Even though such hallucinations rarely happen
in human vision, they are not of a big concern here since, at large
times, the dynamics would be affected by feedback from higher cortical
areas and eye movements, which we are not modeling. Importantly, these
observation suggests that the model performance must be evaluated at
finite, but not asymptotically large times.

\begin{figure}[h]
\centerline{\includegraphics[scale=1.2]{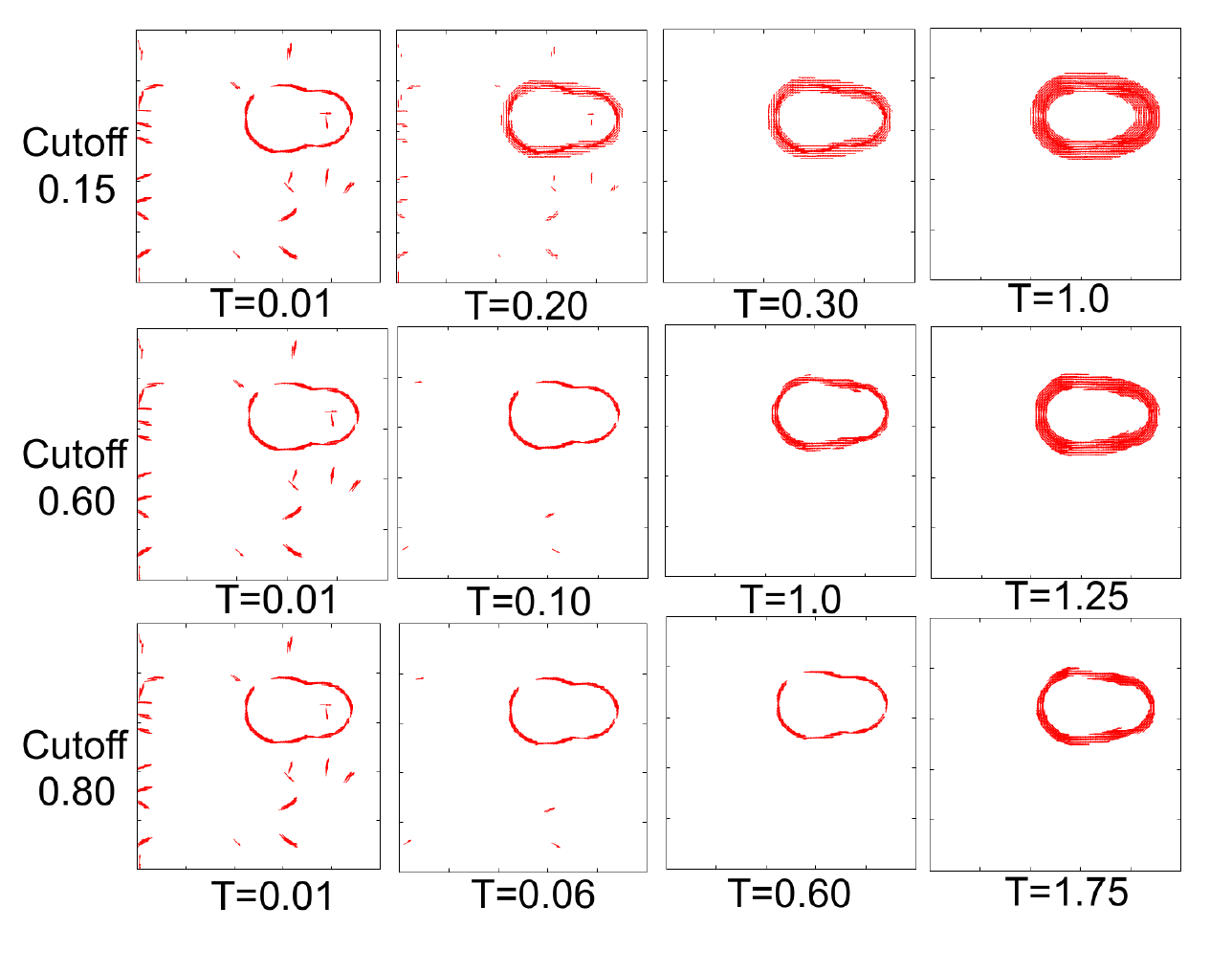}}
\caption{{\bf Neural dynamics at different cutoffs:} Time evolution of a sample image at different cutoff values. At a lower cutoff the occlusions fills rapidly, but it takes longer to suppress the clutter. At higher cutoffs clutter removes quickly, while it takes longer to fill the gaps. Notice the spurious activity around the contours at longer times. This spurious activity is dominant at lower cutoffs.}
 \label{diff_cutoff}
\end{figure}

We quantified the performance in terms of precision, $P$, and recall,
$R$. Precision determines the fraction of the total field activity
integrated over the image that matches the actual target contour
(visible and occluded/invisible). Recall gives the fraction of the
target contour that has been recovered. $P=1$ means that there is no
clutter, and $R=1$ means that all parts of the contour have been
identified. For a successful contour detection, we must have $R,P\to
1$ simultaneously. Both $P$ and $R$ depend on the cutoff used to
decide which neurons are considered active (larger cutoff degrades clutter
faster, but slows down occlusion filling), and on the time of the
simulation (Fig. \ref{diff_cutoff}). Hence different cutoffs and times must be explored.

\begin{figure}[h]
\centerline{\includegraphics[scale=1.4]{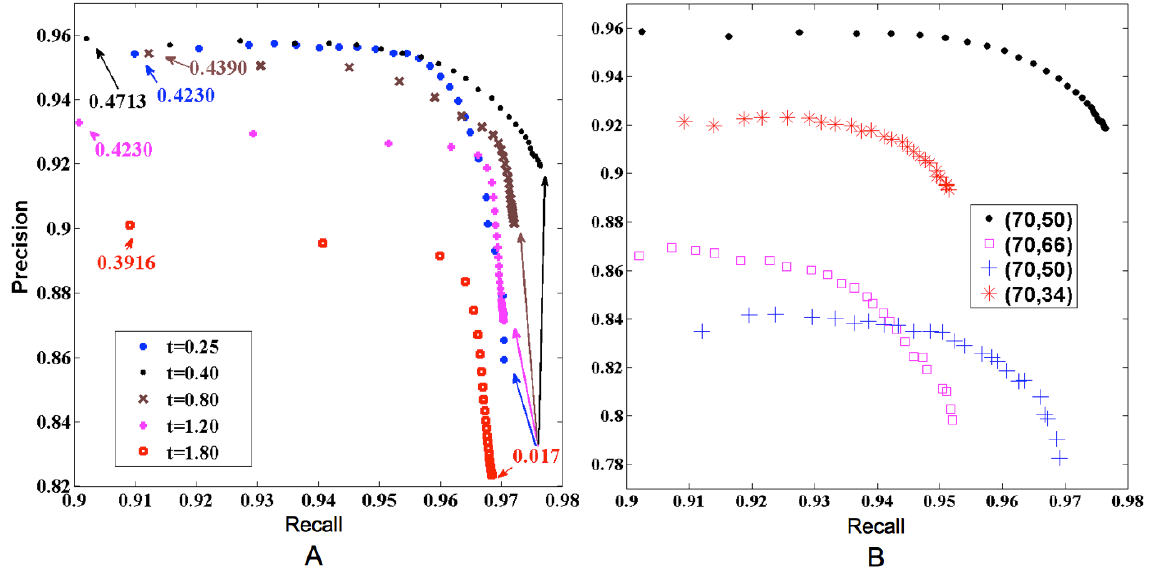}}
 \caption{{\bf Precision vs Recall with an absolute cutoff:} (A) $P$ vs $R$ averaged over 500 randomly generated
     images at various simulation times starting with
     ${(R,P)}=(0.75,0.5)$.  The numbers indicate cutoff values for a
     specific data point at the corresponding simulation time. Note
     the weak dependence on the cutoff. The simulation lengths of
     $t=0.40\times 1/\gamma_{\rm{l}}$ (black dots) produces the curve
     with the best precision and recall combination. 
     (B) $P$ vs
     $R$ with different starting values of precision and recall
     averaged over 100 randomly generated images, but with the same
     model parameters. Legend indicates the initial $(R,P)$. The black
     dots are the same as in the top panel.  Red $*$'s correspond to a
     lower initial precision (more clutter), compared to the black dots. Blue
     $+$'s stand for the same initial $(R,P)$ as black, but with the
     target partitioned into more shorter segments (a larger number of
     occlusions). Pink $\square$'s correspond to higher initial
     precision (less clutter), but the clutter elements are longer and
     harder to suppress. }
 \label{prerec_time}
\end{figure}

Figure \ref{prerec_time}{\bf A} gives the variation of precision and recall
at various cutoffs at particular times during the simulation. At
$t=0$, $(R,P)=(0.75,0.5)$ on average, i.~e., initially about $25\%$ of
the target is invisible and the total lengths of the clutter and the
target segments are nearly equal. At $t$ as small as 0.25 (with
$\Delta t=0.01$), $P,R$ are above 0.9 simultaneously for a large set
of cutoff values ($1\%-42\%$). Since we present the stimulus
instantaneously only, its effect eventually decreases with time. Thus
there is a time that optimizes performance, at which the
precision vs.\ recall curve majorates the same curves for other
times. For the data-set in Fig.~\ref{prerec_time}, this optimal time
is $t=0.40 \times 1/\gamma_{\rm{l}}$ (40 numerical iterations), where
the curve reaches $R\approx 0.97$ and $P\approx 0.95$ simultaneously.

\begin{figure}[h]
\centerline{\includegraphics[scale=1.2]{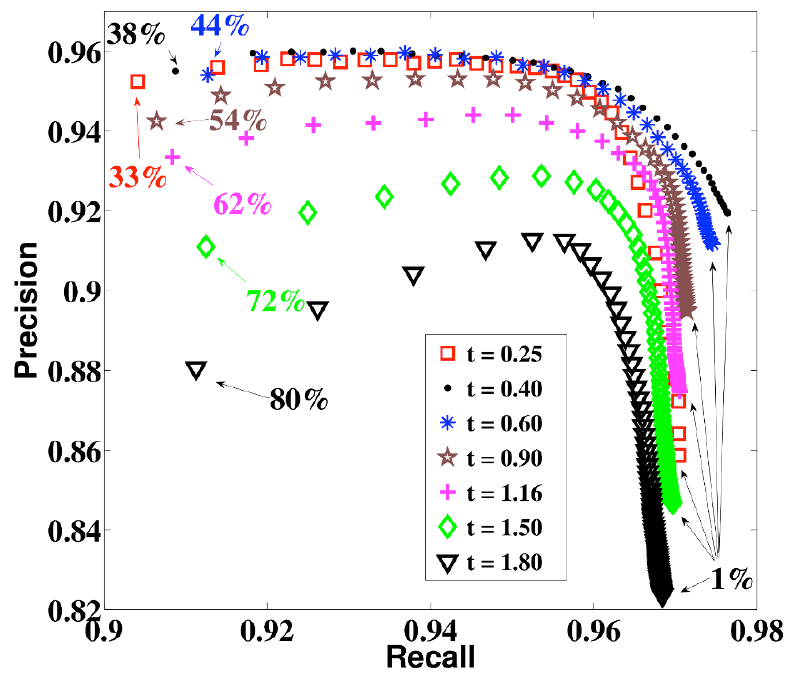}}
\caption{{\bf Precision vs Recall with a relative cutoff:} $P$ vs $R$ averaged over 500 randomly generated
     images at various simulation times starting with
     ${(R,P)}=(0.75,0.5)$.  The numbers indicate cutoff values for a
     specific data point in terms of the percentage of the maximum activity of the field at the corresponding time. Note the similarity with the results in case of absolute cutoff values (Figure 6{\bf{A}}).}
 \label{timeevolution_relative}
\end{figure}

Performance depends only weakly on the {\em ad hoc} details of the
simulations and the data.  For example, defining the threshold
parameter not as an absolute value, but as a fraction of the maximum
activity of the field at a given time point did not change the
precision-recall curves much (Fig.~\ref{timeevolution_relative}).  Similarly, different
amounts of initial clutter had only a moderate effect if the length of
the clutter elements remained the same (Fig.~\ref{prerec_time}{\bf
  B}). This is because the time scale of the clutter decay depends on
the size of the segments, and not on their number. For longer
segments, the decay takes longer, and hence the optimal processing
time increases. The optimal processing time also increases with the
linear dimension of the occlusions present in the target amoebas and
with the number of occlusions (Fig.~\ref{prerec_time}{\bf
  B}). However, for all of these cases, the maximum precision and
recall remain simultaneously high.

\section*{Discussion}
We developed a {\em continuum, coarse-grained model} of V1 to study
contour detection in complex images, which is substantially simpler
than other models in the literature, and yet still performs nontrivial
visual computation.  While borrowing heavily from previous research,
our model differs from most previous approaches by
forgoing individual neurons and describing the neural activity as a
parity-symmetric continuous director field, which makes expressions
for Hebbian connectivity and solutions of the model dynamics
expressible in the closed form.  We incorporate some experimentally
observed properties of the visual neural dynamics, namely non-linear
excitation, thresholding, cross orientation suppression, local
relaxation, global suppression, and, crucially, co-circular excitatory
connectivity \cite{parent1989trace}, which brings long-range context
to local edge detection.

The model identifies long object contours in computer-generated images
with simultaneous recall and precision of over 90\% for many
conditions. It happens even though initially large parts of objects
are invisible (potentially lowering recall), and clutter is present
(decreasing precision). The model fills in the occlusions and filters
out the clutter based on the presence or absence of co-circular
contextual edge support. In addition to the substantial
simplification, this ability to {\em fill in the occlusions}
particularly distinguishes our approach from the previous work on
co-circular excitatory feedback
\cite{parent1989trace,gintautas2011model}. It remains to be seen to
which extent the performance is affected by more natural statistics of
images, and by the presence of stochasticity and synaptic plasticity
in neural dynamics.

The model performs on par or better than agent-based three-dimensional
models (two spatial dimensions and one orientation preference
dimension), with complex, empirically specified co-circular
interaction kernel \cite{gintautas2011model}. This illustrates that
discreteness of neurons, existence of the orientation preference as an
independent variable, and intricate details of the kernel are {\em not
  crucial} for the studied visual processing function.  The reduced
complexity is not only conceptually appealing, but also can result in
more efficient computational implementations. For example, it should
be possible to augment practical feedforward models of object
detection, such as \cite{Serre10042007}, with the laterally connected
layer developed in this work. We expect this to lead to improvements
in object recognition performance.

The model makes predictions that can be tested experimentally, such as
regarding the amount of neural excitation in V1 as a function of the
computation time and the duration of exposure to an image.
Additionally, it predicts that the neural activity localizes to long
contours with time, which can be tested with various imaging
technologies. Finally, it can be used to predict the dependence of the
contour detection performance on the statistical structure of images
and on the exposure time. Testing such predictions in psychophysics
experiments \cite{gintautas2011model} will be a subject of the future
work.

Finally, we notice that the neural field $W(z)=s(x,y)\times
e^{i2\Theta(x,y)}$ can be mapped exactly onto the Landau - de Gennes
order parameter for a two-dimensional nematic liquid crystal \eqqmunu
This may help solve a crucial difficulty in implementing an artificial
laterally-interacting neural model: the computational cost of
long-range communication. Indeed, one can think of materials with
symmetry and dynamical properties such that the neural computation and
the communication are performed by the intrinsic dynamics of the
material itself. Potential implementations can include polarizable
liquid crystals with long-range magnetic interactions, polar colloidal
materials, or heterogenous solid state materials with long-range
connectivity.  The liquid crystal analogy suggests the use of the
well-developed repertoire of theoretical physics to understand the
impact of different terms in the model neural dynamics,
Eq.~(\ref{eq:model}). In particular, one can hope that the future
renormalization group treatment of this dynamics will reveal the terms
in the interaction kernel $K$ that are relevant for its long-time,
long-range aspects.


\section*{Acknowledgments}
We thank G Kenyon, V Gintautas, M Ham, L Bettencourt, and P Goldbart
for stimulating discussions. 


\bibliography{ref}


\newpage

\section*{Figure Legends}

\end{document}